\documentclass[a4paper]{article}

\usepackage{INTERSPEECH2021}
\usepackage{amsmath,graphicx}
\usepackage{subfig}
\usepackage{multirow}
\usepackage{enumitem}
\usepackage{comment}

\title{ Multi-window Data Augmentation Approach for Speech Emotion Recognition }
\name{Sarala Padi$^1$, Dinesh Manocha$^2$, and Ram D.Sriram$^1$}
\address{
  $^1$ITL, NIST, USA    $^2$UMD, College Park, USA}
\email{sarala.padi@nist.gov, dm@cs.umd.edu, ram.sriram@nist.gov}
\begin{document}

\maketitle
\begin{abstract}

We present a  Multi-Window Data Augmentation (MWA-SER) approach for speech emotion recognition. $\textrm{MWA-SER}$ is a unimodal approach that focuses on two key concepts; designing the speech augmentation method and building the deep learning model to recognize the underlying emotion of an audio signal. Our proposed multi-window augmentation approach generates additional data samples from the speech signal by employing multiple window sizes in the audio feature extraction process. We show that our augmentation method, combined with a deep learning model, improves speech emotion recognition performance. We evaluate the performance of our approach on three benchmark datasets: IEMOCAP, SAVEE, and RAVDESS.  We show that the multi-window model improves the SER performance and outperforms a single-window model. The notion of finding the best window size is an essential step in audio feature extraction. We perform extensive experimental evaluations to find the best window choice and explore the windowing effect for SER analysis.
\end{abstract}

\section{Introduction}\label{sec:intro}

In human-computer interactions, the emotion of humans plays a crucial role. It also plays a significant role in psychological and nano-physiological studies of human emotional expression, automatic tutoring systems, customer services, call center services, gaming, personal assistants~\cite{applications}. Thus, there is an increasing demand for automated methods to understand and recognize human emotions~\cite{schuller2018speech,ser-intro-richard}. In real life, there are many ways to express human feelings. Speech is considered an easy and effective communication mechanism to convey human emotions \cite{ser_intro_cowie}. However, automatic speech emotion recognition (SER) is a challenging task~\cite{challenges}, and machine learning models built for such analysis mainly depend on the features extracted from the speech signal. Most of the previous studies on SER obtain spectral and prosody-based features by selecting an appropriate window size~\cite{yoon,tarantino2019self,ramet2018context,neumann2017attentive,savee_Avots}. However, it is crucial to choose an optimal window size, and it depends purely on the type of input signal~\cite{DSP_rabiner,window_zhang}. The features extracted from speech signals vary across humans because of speaking style, native language speaker status, speaking speed, language usage, context, location. Also, there are significant intercultural differences in understanding human emotions~\cite{influence}. Thus, it is hard to build speaker-independent models and quantify emotions for a given sentence. 

In recent years,  deep learning (DL) models have shown great success in recognizing the emotions from the manually obtained features or by directly extracting the speech characteristics from the raw waveform or spectrograms~\cite{yoon,lee2015high, mirsamadi, DBNs,kim2019dnn,han2014speech,sarma2018emotion,neumann2017attentive, CNN-adie-schuller,ramet2018context,tarantino2019self, 
satt2017efficient,han2014speech,CNN-RNN-schuller}. However, building such complex models requires large amounts of data to learn the millions of parameters. Also, overfitting is one of the challenges in building such models. In overcoming this issue, there are some techniques incorporated during training, such as making use of dropout layers, adding regularizers, normalizing each layer input by adding batch normalization, incorporating transfer learning techniques, and making use of pre-trained models~\cite{tl_survey}. In contrast to these techniques, data augmentation is another approach that overcomes model overfitting by providing more generalized data during training.  However, there are few methods explored, such as vocal tract length perturbations (VTLP), signal-based transformations like time stretch, pitch shift, and adding noise to the original utterance, Generative Adversarial Networks (GAN), and CycleGAN augmentation approaches for SER tasks~\cite{speech_aug, speech_aug_gan,sarma2018emotion, speech_aug_gan,speech_aug_cygan,gan_sahu}. The main disadvantage of signal-based transformations is that models tend to overfit due to similar samples in the training set, while random balance removes possibly valuable information. The downside with GAN and CycleGAN models is that the feature vectors generated from these models dependent on data used during training and may not generalize to other datasets. The other big concern with these models is that they are hard to train and optimize. 

We propose a Multi-window Data Augmentation (MWA-SER) approach to address one of the challenges (overfitting) in building the DL models for SER analysis. The proposed method generates more data samples from the speech signal by employing multiple window sizes in the audio feature extraction process. The advantage of our approach is twofold:  by applying the multi-window approach, we can obtain features considering smaller and longer utterances, which play a vital role in emotion analysis, and 2) address the lack of data to build a complex deep learning model by generating additional data samples from the speech signal. To show the benefit of our proposed approach, we consider three benchmark datasets: IEMOCAP, SAVEE, and RAVDESS, and show that our proposed strategy improves the emotion recognition performance and outperforms the single-window model. When we work with a multi-window approach, finding the best window size is an essential step in audio feature extraction. We also extend our study to explore the windowing effect and to find the best window choice for speech emotion recognition analysis.


\begin{figure*}[ht!]
\begin{center}
 \includegraphics[width=16cm,height=6cm]{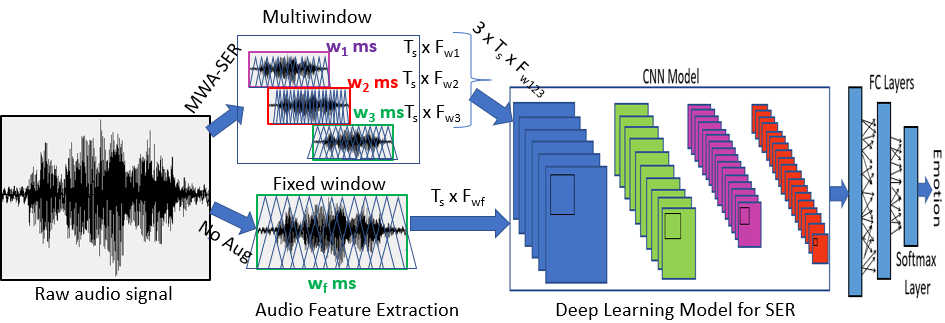}
\end{center}
\vspace{-0.4cm}
   \caption{A multi-window data augmentation combined with a deep learning model for SER analysis. $W_{fs}$ indicates fixed window model, $T_{s}$ means the number of examples used for training, $F$ indicates features extracted using a fixed or a multi-window.   }
\label{fig:method}
\end{figure*}

\section{MWA-SER: Methodology}\label{sec:method}

The proposed methodology consists of two stages: 1) a multi-window data augmentation and 2) a Convolutional Neural Network (CNN) model. The first stage generates additional data samples to augment the CNN model by employing multiple window sizes in the audio feature extraction process. In the second stage, the CNN model uses the extracted features for emotion analysis. As shown in Figure~\ref{fig:method}, in the audio feature extraction process, a $\textrm{25~ms}$ window has densely placed filters, and a $\textrm{200~ms}$ window has filters that are farther apart. We can also see that windows are placed at different scales to extract features from smaller and longer utterances. A single-window model (with ``$T_{s}$'' number of training examples) extracts ``$T_{s} \times {\textrm{Features}}$''  during training. On the other hand, a multi-window model (with three window sizes) extracts ``$3\times T_{s} \times {\textrm{Features}}$'' , which indeed increases the number of examples used for training from ``$1 \times T_{s}$'' to $``3 \times T_{s}$''.  

\subsection{Stage 1: Multi-window Approach}

In speech processing,  features extracted using a windowed signal which is typically called a frame.  In general, an audio signal is stationary for a short period, and the width of the windowing function has a direct impact on the feature values extracted from the speech signal~\cite{window_kelly2011effects}.  For example, if the selected window length\footnote{The number of samples considered for short-time analysis} is too small,  then we may not have enough data samples to get a reliable spectral estimate of the signal; if it is too large, then the signal changes too much throughout the frame. So, selecting the width of a windowing function is a crucial step, and it is hard to choose when no background information about the input signal is known~\cite{DSP_rabiner,window_zhang}.   Studies have shown that an optimal window size selection increases the correlation between the acoustic representation and human perception of a speech signal~\cite{window_stft_nisar,window_kirkpatrick}. 

A window function is represented by three tuples: window width (milliseconds), window offset or overlap (milliseconds), and window shape. To extract part of a signal, we multiply the value of the signal at time ``t'', signal[t], with the value of the window (hamming or rectangular) at time ``t'', window[t]  represented as:
\begin{equation}
windowed_{signal[t]}= window[t] \times signal[t].
\end{equation}
Windowed signal (frame) is used to compute features for emotion analysis. For SER, a typical window of size $\textrm{25~ms}$ is used to extract features with an overlap of $\textrm{10~ms}$ window~\cite{yoon,tarantino2019self,ramet2018context,neumann2017attentive}.  On the other hand,  other studies have shown that a  bigger window size increases the emotion recognition performance\cite{chernykh2017emotion,tripathi}.  There are other studies in evaluating the importance of step size (overlap window size), but a single-window is used for SER analysis~\cite{tarantino2019self,chernykh2017emotion}. Tarantino et al. demonstrated the effect of overlap window size for SER  and showed that a small step size leads to a reduced test loss.  Chernykh et al. experimented with different window sizes ranging from $\textrm{30~ms}$ to $\textrm{200~ms}$ and selected a single window of size $\textrm{200~ms}$ for SER analysis. To the best of our knowledge, our approach is the first method to generate additional data samples to augment the  DL models by employing multiple window sizes in the audio feature extraction process along with a CNN model for SER analysis. 


\subsection{Stage 2: Convolutional Neural Network Model (CNN)}\label{subsec:cnn}

Motivated by the CNN model, which outperformed ML models for the SER task~\cite{venkat_Dl_outper_ml}, we build a CNN model to recognize the underlying emotion of the speech signal. We use four convolutional layers in the CNN model with two fully-connected (FC) layers followed by a softmax layer. The number of kernels used in each of the four convolutional layers is $32$, $64$, $128$, $256$ followed by $128$ and $32$ hidden neurons at the FC layers.  To mitigate the overfitting issue, we use dropout layers with $\textrm{p}=0.23$.  Furthermore, after each convolutional layer, we used batch normalization and max-pooling layers to subsample the feature dimensions. We use kernels of different sizes to learn the representative features to a larger context. The typical kernel sizes that we use are $7 \times 7$, $5 \times 5$, $3 \times 3$,  and $1 \times 1$.  During training, we augment the CNN model by generating additional samples using multiple windows. All the extracted features are fed to the CNN model to learn the most representative features for SER analysis. Finally, the softmax layer is applied, to the vector generated by the FC layer, to predict the emotion.

\subsection{Features:}
As detailed in~\cite{chernykh2017emotion,tripathi}, before feature extraction, the speech signal is preprocessed with methods such as DC removal and then normalized with the maximum value of the signal. We extract 34-dimensional features: $26$ spectral features: $13$-dimensional Mel Frequency Cepstral Coefficients (MFCC), $13$-dimensional Chroma-based features, and $8$ time-domain features; zero-crossing rate, spectral flux, spectral centroid, spectral entropy, spectral rolloff, short-term energy, and short term entropy of energy.  We extract speech features for the first $200$ frames. Longer signals cut down to $200$ frames, and shorter signals are padded with zeros.


\section{Experiments}\label{sec:exp}
\subsection{Datasets:}\label{subsec:data}
To evaluate and compare our proposed methodology with the baseline, we conduct the experiments on three benchmark datasets.\\\\
{\bf IEMOCAP}~\cite{iemocap}: Interactive emotional dyadic motion capture (IEMOCAP) dataset contains improvised and scripted multimodal dyadic conversations between actors of the opposite gender. It consists of 12 hours of speech data from 10 subjects. It includes nine categorical emotions and 3-dimensional labels. We consider categorical emotions, where at least two experts agree in the annotation.  We conduct two experiments: 1) Exp~1  uses four classes of emotions: ``angry'', ``happy'', ``neutral'', ``sad'', and 2) Exp~2 uses the same categories as in Exp~1, replacing the ``happy'' emotion with ``excited''.  The total number of examples used for Exp~1 is $4490$, and the number of examples per category is $1103$, $595$, $1708$, and $1084$, respectively. The number of examples in the ``excited'' category is $1041$, making the total number of examples used for Exp~2  $4936$.\\\\
{\bf ~RAVDESS}~\cite{ravdess} : The Ryerson Audio-Visual Data of Emotional Speech and Song ($\textrm{RAVDESS}$) is a standard scripted speech dataset containing eight categories of emotions.  This dataset contains 60 spoken sentences and 40 sung sentences  from 24 actors (12 male, 12 female). We consider only speech sentences from 24 actors for emotion analysis. We merged ``neutral'' and ``calm'' emotions into a single class called ``neutral''. We  consider six categories of emotions:``angry'', ``happy'', ``neutral'', ``sad'', ``fear'', and ``disgust'' for  SER analysis. \\\\
{\bf ~SAVEE}~\cite{savee} : The Surrey Audio-Visual Expressed Emotion ($\textrm{SAVEE}$) dataset has 120 spoken sentences from 4 native English male actors. We consider six categories of emotions:``angry'', ``happy'', ``neutral'', ``sad'', ``fear'', and ``disgust'' for  SER analysis.\\

\subsection{Experimental Setup and Training Details:}

We use the scipy library\footnote{www.scipy.org} to extract speech-based features and the Keras deep learning framework with TensorFlow 2.0 as the backend to build the CNN model. We train the  CNN model with an RMSprop optimizer to minimize the categorical cross-entropy loss with a $0.0004$ learning rate.  We ran the model for $1000$ epochs with a batch size of $32$.  In our evaluations, we use $80\%$ of the data for training and $20\%$ of data for testing purposes because performing cross-validation on deep learning models with varying window sizes is not feasible in terms of time and computational requirements. We present our findings by reporting Unweighted Accuracy (UA), Weighted Average Precision (WAP), and Weighted Average F1 (WAF1) measures. 

In the experiments, the window sizes that we considered to evaluate the performance of single-window and multi-window approaches are $\textrm{25~ms}$, $\textrm{50~ms}$, $\textrm{100~ms}$, and $\textrm{200~ms}$ with a $50\%$ overlap.  In the multi-window approach,  we train the DL  model on the data samples generated by multiple-window sizes. For example, in a three-window model with $\textrm{25~ms}$, $\textrm{50~ms}$ and  $\textrm{100~ms}$ window sizes, we used all three windows to generate data for training, and while testing, we considered the window that gives the best accuracy against the test data.

\begin{table}[hbt!]
\caption{Performance comparison between  a single-window model and a multi-window methods on the $\textrm{IEMOCAP}$, $\textrm{SAVEE}$, and $\textrm{RAVDESS}$ datasets.  The single-window model with  $\textrm{25~ms}$ window size is a baseline. For the multi-window approach, we reported the best window model performances.  Abbreviations: A-Angry, H- Happy, N-Neutral, E-Excited.   }
\label{tab:Comp_2}
\renewcommand{\arraystretch}{1.2}
\resizebox{0.45\textwidth}{!}{
\centering
\begin{tabular}{ c|l|c|c|c} 
 \hline
{Dataset}                            &{Model}                                &{UA (\%)} &{WAP (\%)}& WAF1 (\%)  \\\hline
                                           
\multirow{2}{*}{\parbox{3cm}{IEMOCAP: A, H, S, N \\ (Exp 1) }} & Single-window                           &60     &65    &61\\
                           & {\bf MWA-SER}                         &{\bf 65}   &{\bf 73} &{\bf 68} \\\hline               
                                      
\multirow{2}{*}{\parbox{3cm}{IEMOCAP: A, E, S, N \\ (Exp 2)}} &Single window                          &60  &64 &61    \\
                              & {\bf MWA-SER}                        &{\bf 66}   &{\bf 68}  &{\bf 66}\\\hline         
                                      
    \multirow{2}{*}{SAVEE}          &Single-window                 &56           &67& 59\\
                                   &{\bf MWA-SER }                &{\bf 70}     &{\bf 74} &{\bf 71}\\\hline
    \multirow{2}{*}{RAVDESS}         &Single-window               &86            &86  &86\\
                                 &{\bf MWA-SER }                 &{\bf 88}      &{\bf 88}   &{\bf 88}\\\hline
    \end{tabular}} 

\end{table}
\begin{figure*}[htb!]
\begin{center}
 \subfloat[\small {Exp 1:A,H,N,S}]{\includegraphics[width=4.2cm,height=4.2cm]{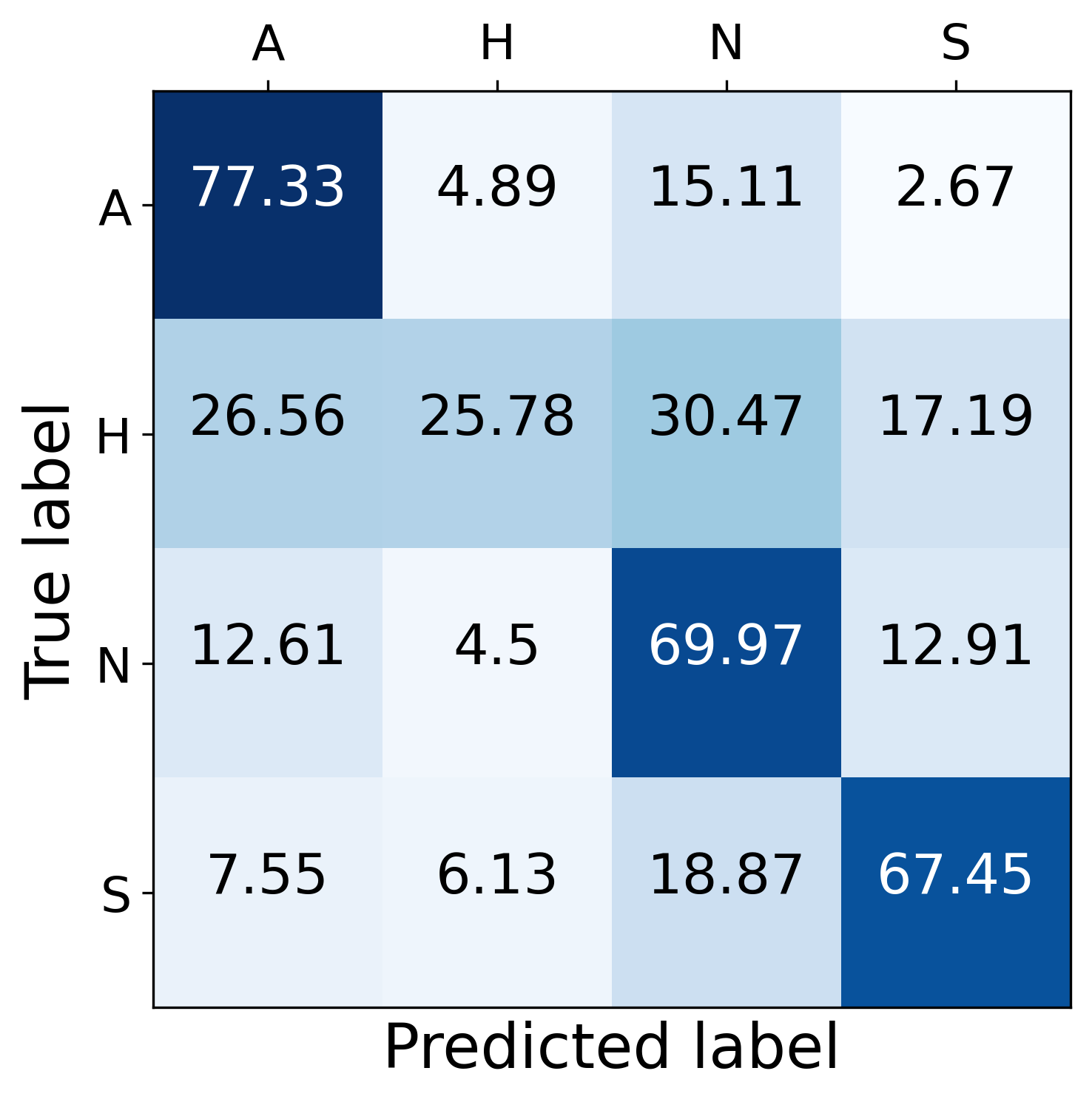}}
 \subfloat[\small{Exp 2: A,E,N,S}]{\includegraphics[trim=1cm 0cm 0cm 0cm, clip=true, width=4.2cm,height=4.2cm]{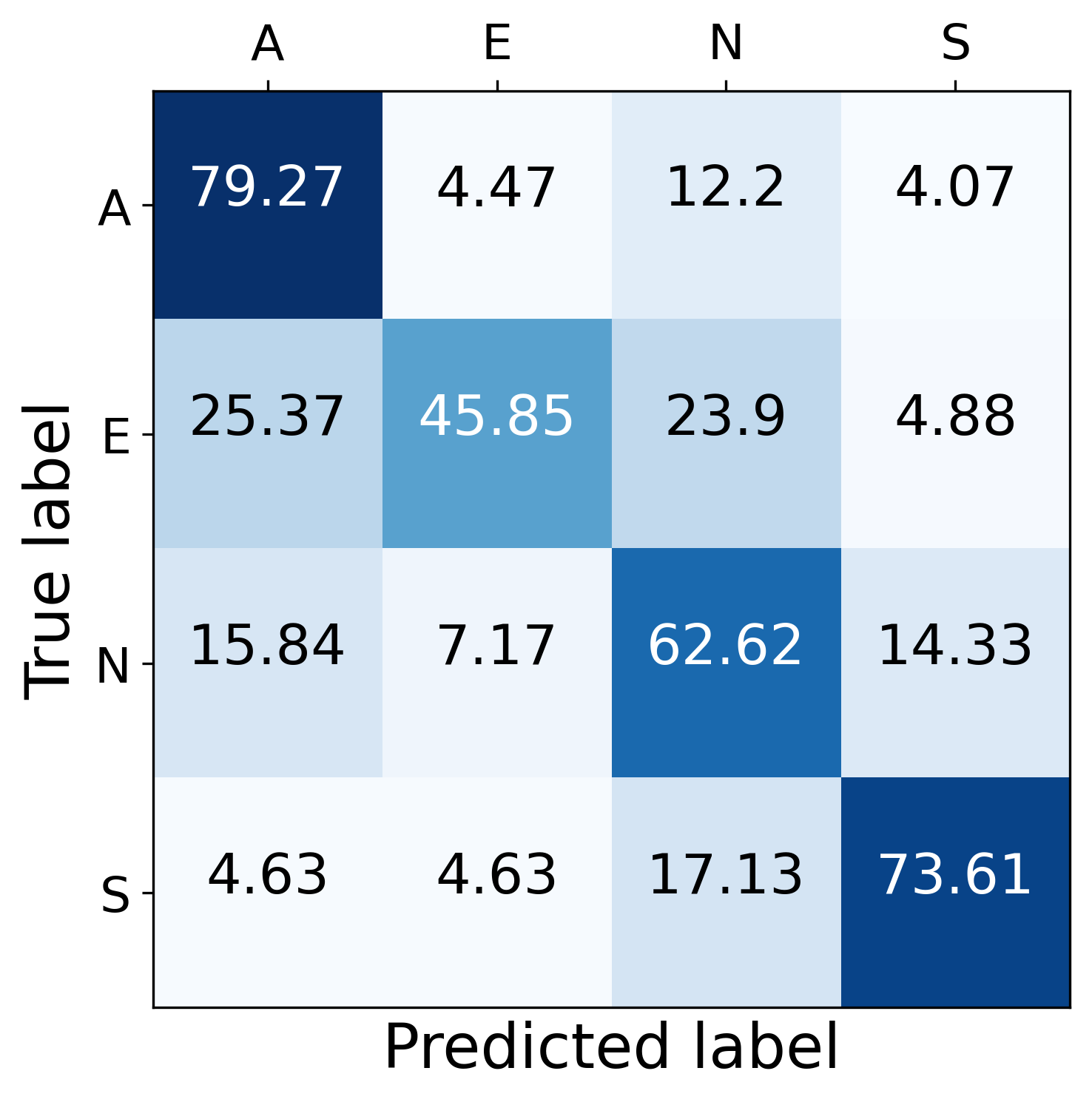}}
 \subfloat[\small{SAVEE}]{\includegraphics[trim=1cm 0cm 0cm 0cm, clip=true, width=4.2cm,height=4.2cm]{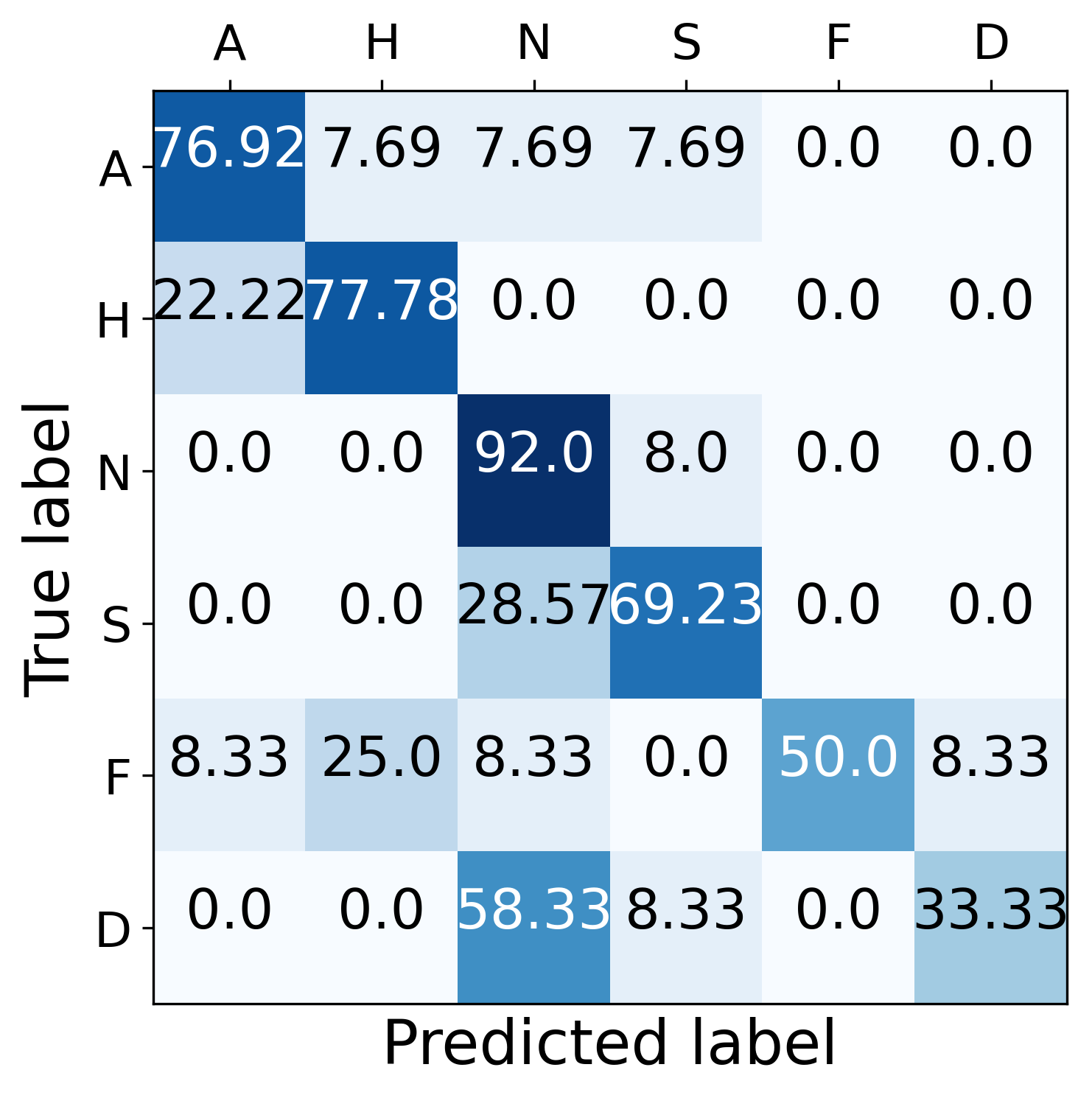}}
 \subfloat[\small{RAVDESS}]{\includegraphics[ trim=1cm 0cm 0cm 0cm, clip=true, width=4.2cm,height=4.2cm, width=4.2cm,height=4.2cm]{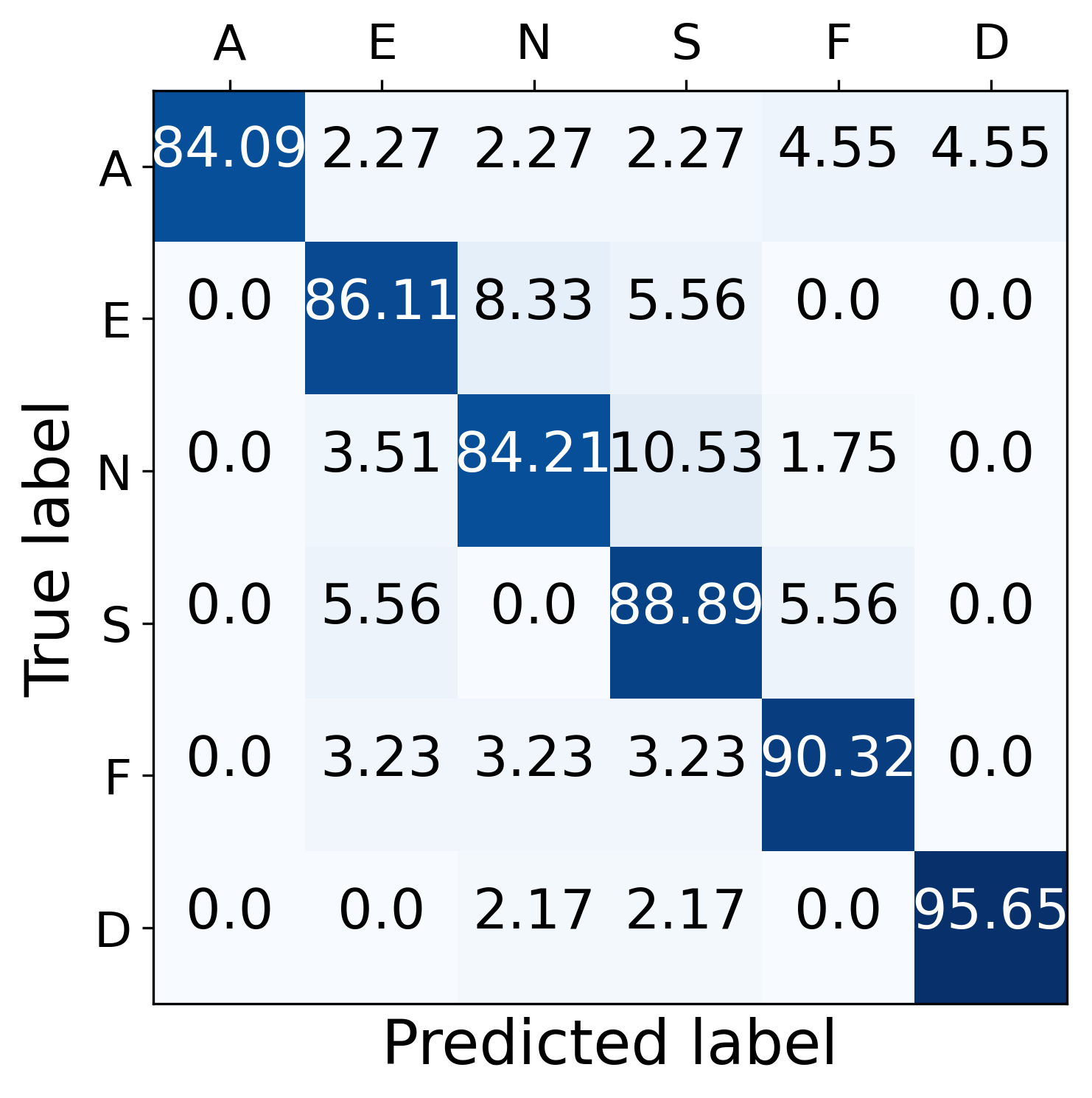}}
\end{center}
   \caption{ Confusion matrix of the proposed MWA-SER approach evaluated for IEMOCAP (Exp 1 and 2), SAVEE, and RAVDESS datasets.  Abbreviations: A-angry, H-happy, N-neutral, E-excited,  F-Fear, D-Disgust. }
\label{fig:confusion}
\end{figure*}

\begin{table*}[hpbt!]
\caption{ The performance comparison of a single-window and multi-window models by varying the window size and increasing the data to augment the DL models for SER analysis.  The number of windows that we experimented with  are  $\textrm{25~ms} (w1)$, $\textrm{50~ms} (w2)$, $\textrm{100~ms} (w3)$ and $\textrm{200~ms} (w4)$. w12 indicates a two window model with  window sizes $\textrm {25~ms}$ and $\textrm{50~ms}$. This model generates $2\times T_{s}$ data examples. Similarly, w13, w123, and w1234 indicate two, three, and four windows used to generate additional examples to augment the DL model for SER analysis. The bold letters indicate the optimal window choice for the $\textrm{IEMOCAP}$, $\textrm{SAVEE}$, and $\textrm{RAVDESS}$ datasets.   }\label{tab:window_sr}
\renewcommand{\arraystretch}{1.15}
\resizebox{0.99\textwidth}{!}{
\begin{tabular}{|c|c|c|c|c|c|c|c|c|c|c|c|c|c|c|c|c|c|} 
 \hline
\multirow{3}{*}{\bf {Dataset}} &\multirow{3}{*}{ {\bf Metric}} &\multicolumn{4}{|c|}{\multirow{2}{*}{\bf{Single window}}} &\multicolumn{11}{|c|}{\bf{Multi-window}}\\\cline{7-17}
                           &      &\multicolumn{4}{|c|}{}           &\multicolumn{6}{|c|}{Two window} &\multicolumn{4}{|c|}{Three window} &\multicolumn{1}{|c|}{Four window} \\\cline{3-17}
                          &     &25(w1) &50(w2)&100(w3)&200(w4)         &w12 &w13  &w14  &w23  &w24  &w34       & w123 &w124  &w134  &w234  &w1234\\\hline
                            &Accuracy   &60   &62   & 60  &61      &64   &63   &62   &64   &63   &63       &64   & 65  &65   &{\bf 65}   &64\\\cline{3-17}
  \multirow{3}{*}{IEMOCAP (Exp 1) }&WAP       &65   &67   &66   &72      &70   &67   &67   &71   &72   &70       &72   &68   &71   &{\bf 73}   &70\\\cline{3-17}
                    &WAF1    &61   &64   &62   &66       &66   &64   &64   &66   &66   &65       &67   &66   &67   &{\bf 68}    &66\\\hline
                 &Accuracy   &60   &63   &62   &62         &65   &62   &62   &64   &61   &62       &{\bf 66}   & 65  & 65  & 64  &66\\\cline{3-17}
  \multirow{3} {*}{IEMOCAP (Exp 2)} &WAP&64   &65   &66   &64         & 67  & 63  &64   &66   & 61  & 63      &{\bf 68}   &67   &66   &67   &67\\\cline{3-17}
                           &WAF1     &61   &64   &63   &62         &66   &62   & 63  &64   & 61  &62       &{\bf 66}   & 65  &65   &65    &66\\\hline
                    &Accuracy     &56   &62  &67    &61          &64    &65   &69   &65   &69   &{\bf 70}     &64   &68   &65   &67    &70\\\cline{3-17}
\multirow{3}{*}{SAVEE} &WAP           &67   &68  &69    &66         &69    &72   &71   &73   &72   &{\bf 74}     &72   &71   &72   &68    &73   \\\cline{3-17}
           &WAF1                &59   &63   &67   &62         &65    &67   &69   &67   &70   &{\bf 71}     &66   &68   &67   &67    &70  \\\hline
         &Accuracy              &86   &82   &82   &83         &84    &85   &83   &86   &82   &83    &85    &84   &85   &85  &{\bf 88} \\\cline{3-17}
\multirow{3}{*}{RAVDNESS} &WAP        &86   &82   &84   &84          &86    &86   &84   &87   &82   &84    &86    &85   &86   &86  &{\bf 88}\\\cline{3-17}
           &WAF1                        &86   &82   &83   &83          &84    &85   &83    &86  &82   &83    &85    &83   &85   &85  &{\bf 88}  \\\hline
\end{tabular}}

\end{table*} 
\section{Experimental Results:}\label{sec:results}

Table~\ref{tab:Comp_2} shows an SER performance comparison of single-window and multi-window methods on the $\textrm{IEMOCAP}$, $\textrm{SAVEE}$ and $\textrm{RAVDESS}$ datasets. For the IEMOCAP dataset ($\textrm{Exp~1}$), our proposed MWA-SER approach improved the emotion recognition accuracy with an improvement of $6\%$ accuracy,  $9\%$ in WAP, and $7\%$ in WAF1 measure.  We also extended our analysis by replacing ``happy'' emotion with ``excited'' and showed that the proposed model outperformed a single-window model with an improvement of $6\%$ accuracy,  a $7\%$ in WAP, and a $5\%$ in WAF1, respectively. We also investigate the performance of our proposed augmentation approach on two other benchmark datasets: $\textrm{SAVEE}$ and $\textrm{RAVDESS}$, and Table~\ref{tab:Comp_2} compare the performance of single-window and multi-window models on these two datasets.  For the $\textrm{SAVEE}$ dataset, there is an improvement of $14\%$  accuracy, $7\%$ in WAP, and $12\%$ in WAF1 score. For the $\textrm{RAVDESS}$ dataset, there is an improvement of $2\%$  accuracy, WAP, and  WAF1 scores.

From experimental analysis, we observed that tuning CNN model parameters are hard to classify the ``happy'' class examples than the other three categories of emotions. The potential reason is some examples belonging to the ``happy'' category include overlapped speech. Further, the duration of the ``happy'' category examples is small compared to other emotional classes. It may not be accurate for all the examples of the ``happy'' emotion in the dataset. 

To further analyze the misclassification error within and across emotional categories, we show the confusion matrices for the three benchmark datasets. Figure~\ref{fig:confusion}(a) and (b)  show the confusion matrices for IEMOCAP dataset (Exp~1 and 2).  The ``happy'' and ``excited'' classes of emotions got confused with the ``neutral'' class of emotion and got the best accuracy for the other three categories of emotions.  Similarly, for the $\textrm{SAVEE}$ and $\textrm{RAVDESS}$ datasets (Figure~\ref{fig:confusion}(c) and (d)), ``sad'' and ``neutral'' emotions got confused and in fact these are the most confusable emotions across three datasets. 

\section{Windowing Effect:}

To further study the importance of window size, we extended our analysis by varying the window size and increasing the data to augment the DL model for SER analysis. Table~\ref{tab:window_sr} shows the performance comparison of a single-window and multi-window models. As shown in Table~\ref{tab:window_sr}, for  $\textrm{IEMOCAP}$ dataset the three-window model with window sizes $50,100,200$~ms for Exp~1 and $25,50,100$~ms for Exp~2 performs better than  single-window model. It indicates that extraction of audio features at multiple scales is useful for emotion analysis. However, the optimal window size that we use in our experimental study may vary depending on the dataset because our methodology purely depends on the features extracted from the speech signal. These extracted features vary widely across datasets based on the type of audio signal,  duration of an audio signal, etc. For instance, we extended our experiments on two other benchmark datasets: SAVEE and RAVDESS.  As shown in Table~\ref{tab:window_sr},  the  $\textrm{SAVEE}$ dataset, the two-window model with $100~ms$ and $200~ ms$ window sizes, and for the $\textrm{RAVDESS}$ dataset, a four window model with $25~ms$, $50~ms$, $100~ms$, and $200~ ms$ window sizes improved the SER performance.  The performance of the proposed model and the optimal window choice depends on: the type of dataset, type of features extracted, type of emotions considered for analysis, window size used to compute the speech-based features, and the model architecture used for the SER analysis. However, this general idea can be adapted to speech processing applications when limited data are available to train complex DL models.    

\section{Conclusions and Future Work}\label{sec:con}
We presented a multi-window data augmentation approach to improving speech emotion recognition performance. We addressed the overfitting issue of the deep learning model by generating additional data samples during training.   We evaluated the performance of the proposed approach on three benchmark datasets.  We showed that our proposed method improved the emotion recognition performance and consistently outperformed the single-window model across datasets.  We also showed that window size plays a crucial role in speech emotion recognition, and the best window choice varies across datasets.  In the future, to further improve the emotion recognition performance, we plan to extend the multi-window approach with other augmentation techniques such as spectrogram and GAN-based methods.  

{\bf \large Disclaimer:} The views and conclusions presented in this paper are those of the authors and should not be interpreted as the official findings, either expressed or implied, of NIST or the U.S. Government.

\bibliographystyle{IEEEtran}

\bibliography{Padi}

\begin{thebibliography}{10}
\providecommand{\url}[1]{#1}
\csname url@samestyle\endcsname
\providecommand{\newblock}{\relax}
\providecommand{\bibinfo}[2]{#2}
\providecommand{\BIBentrySTDinterwordspacing}{\spaceskip=0pt\relax}
\providecommand{\BIBentryALTinterwordstretchfactor}{4}
\providecommand{\BIBentryALTinterwordspacing}{\spaceskip=\fontdimen2\font plus
\BIBentryALTinterwordstretchfactor\fontdimen3\font minus
  \fontdimen4\font\relax}
\providecommand{\BIBforeignlanguage}[2]{{%
\expandafter\ifx\csname l@#1\endcsname\relax
\typeout{** WARNING: IEEEtran.bst: No hyphenation pattern has been}%
\typeout{** loaded for the language `#1'. Using the pattern for}%
\typeout{** the default language instead.}%
\else
\language=\csname l@#1\endcsname
\fi
#2}}
\providecommand{\BIBdecl}{\relax}
\BIBdecl

\bibitem{applications}
A.~Ko{\l}akowska, A.~Landowska, M.~Szwoch, W.~Szwoch, and M.~R. Wrobel,
  ``Emotion recognition and its applications,'' in \emph{Human-Computer Systems
  Interaction: Backgrounds and Applications 3}.\hskip 1em plus 0.5em minus
  0.4em\relax Springer, 2014, pp. 51--62.

\bibitem{schuller2018speech}
B.~W. Schuller, ``Speech emotion recognition: Two decades in a nutshell,
  benchmarks, and ongoing trends,'' \emph{Communications of the ACM}, vol.~61,
  no.~5, pp. 90--99, 2018.

\bibitem{ser-intro-richard}
H.~Richard, R.~Tom, R.~Yvonne, and S.~Abigail, \emph{Being Human:
  Human-Computer Interaction in The Year 2020}.\hskip 1em plus 0.5em minus
  0.4em\relax Report, Microsoft Corporation, 2008.

\bibitem{ser_intro_cowie}
R.~Cowie, E.~Douglas-Cowie, N.~Tsapatsoulis, G.~Votsis, S.~Kollias, W.~Fellenz,
  and J.~G. Taylor, ``Emotion recognition in human-computer interaction,''
  \emph{IEEE Signal Processing Magazine}, vol.~18, no.~1, pp. 32--80, 2001.

\bibitem{challenges}
L.~Devillers, L.~Vidrascu, and L.~Lamel, ``Challenges in real-life emotion
  annotation and machine learning based detection,'' \emph{Neural Networks},
  vol.~18, no.~4, pp. 407--422, 2005.

\bibitem{yoon}
S.~Yoon, S.~Byun, and K.~Jung, ``Multimodal speech emotion recognition using
  audio and text,'' in \emph{2018 IEEE Spoken Language Technology Workshop
  (SLT)}.\hskip 1em plus 0.5em minus 0.4em\relax IEEE, 2018, pp. 112--118.

\bibitem{tarantino2019self}
L.~Tarantino, P.~N. Garner, and A.~Lazaridis, ``Self-attention for speech
  emotion recognition,'' \emph{Proc. Interspeech 2019}, pp. 2578--2582, 2019.

\bibitem{ramet2018context}
G.~Ramet, P.~N. Garner, M.~Baeriswyl, and A.~Lazaridis, ``Context-aware
  attention mechanism for speech emotion recognition,'' in \emph{2018 IEEE
  Spoken Language Technology Workshop (SLT)}.\hskip 1em plus 0.5em minus
  0.4em\relax IEEE, 2018, pp. 126--131.

\bibitem{neumann2017attentive}
M.~Neumann and N.~T. Vu, ``Attentive convolutional neural network based speech
  emotion recognition: A study on the impact of input features, signal length,
  and acted speech,'' \emph{arXiv preprint arXiv:1706.00612}, 2017.

\bibitem{savee_Avots}
E.~Avots, T.~Sapinski, M.~Bachmann, and D.~Kaminska, ``Audiovisual emotion
  recognition in wild,'' \emph{Machine Vision and Applications}, vol.~30, pp.
  975--985, 2018.

\bibitem{DSP_rabiner}
L.~R. Rabiner and R.~W. Schafer, \emph{Introduction to digital speech
  processing}.\hskip 1em plus 0.5em minus 0.4em\relax Now Publishers Inc, 2007.

\bibitem{window_zhang}
S.~Zhang, E.~Loweimi, P.~Bell, and S.~Renals, ``Windowed attention mechanisms
  for speech recognition,'' in \emph{ICASSP}.\hskip 1em plus 0.5em minus
  0.4em\relax IEEE, 2019, pp. 7100--7104.

\bibitem{influence}
R.~Altrov and H.~Pajupuu, ``The influence of language and culture on the
  understanding of vocal emotions,'' \emph{Eesti ja soome-ugri keeleteaduse
  ajakiri. Journal of Estonian and Finno-Ugric Linguistics}, vol.~6, no.~3, pp.
  11--48, 2015.

\bibitem{lee2015high}
J.~Lee and I.~Tashev, ``High-level feature representation using recurrent
  neural network for speech emotion recognition,'' in \emph{Sixteenth annual
  conference of the international speech communication association}, 2015.

\bibitem{mirsamadi}
S.~Mirsamadi, E.~Barsoum, and C.~Zhang, ``Automatic speech emotion recognition
  using recurrent neural networks with local attention,'' in
  \emph{ICASSP}.\hskip 1em plus 0.5em minus 0.4em\relax IEEE, 2017, pp.
  2227--2231.

\bibitem{DBNs}
\BIBentryALTinterwordspacing
S.~Latif, R.~Rana, S.~Younis, J.~Qadir, and J.~Epps, ``Transfer learning for
  improving speech emotion classification accuracy,'' in \emph{Proc.
  Interspeech 2018}, 2018, pp. 257--261. [Online]. Available:
  \url{http://dx.doi.org/10.21437/Interspeech.2018-1625}
\BIBentrySTDinterwordspacing

\bibitem{kim2019dnn}
E.~Kim and J.~W. Shin, ``Dnn-based emotion recognition based on bottleneck
  acoustic features and lexical features,'' in \emph{ICASSP}.\hskip 1em plus
  0.5em minus 0.4em\relax IEEE, 2019, pp. 6720--6724.

\bibitem{han2014speech}
K.~Han, D.~Yu, and I.~Tashev, ``Speech emotion recognition using deep neural
  network and extreme learning machine,'' in \emph{Fifteenth annual conference
  of the international speech communication association}, 2014.

\bibitem{sarma2018emotion}
M.~Sarma, P.~Ghahremani, D.~Povey, N.~K. Goel, K.~K. Sarma, and N.~Dehak,
  ``Emotion identification from raw speech signals using {DNNs},'' in
  \emph{Proc. INTERSPEECH}, 2018, pp. 3097--3101.

\bibitem{CNN-adie-schuller}
G.~Trigeorgis, F.~Ringeval, R.~Brueckner, E.~Marchi, M.~A. Nicolaou,
  B.~Schuller, and S.~Zafeiriou, ``Adieu features? end-to-end speech emotion
  recognition using a deep convolutional recurrent network,'' in \emph{2016
  IEEE international conference on acoustics, speech and signal processing
  (ICASSP)}.\hskip 1em plus 0.5em minus 0.4em\relax IEEE, 2016, pp. 5200--5204.

\bibitem{satt2017efficient}
A.~Satt, S.~Rozenberg, and R.~Hoory, ``Efficient emotion recognition from
  speech using deep learning on spectrograms.'' in \emph{INTERSPEECH}, 2017,
  pp. 1089--1093.

\bibitem{CNN-RNN-schuller}
G.~Keren and B.~Schuller, ``Convolutional {RNN}: an enhanced model for
  extracting features from sequential data,'' in \emph{Proc. IEEE International
  Joint Conference on Neural Networks (IJCNN)}, 2016, pp. 3412--3419.

\bibitem{tl_survey}
K.~Weiss, T.~M. Khoshgoftaar, and D.~Wang, ``A survey of transfer learning,''
  \emph{Journal of Big data}, vol.~3, no.~1, p.~9, 2016.

\bibitem{speech_aug}
C.~Etienne, G.~Fidanza, A.~Petrovskii, L.~Devillers, and B.~Schmauch, ``Cnn+
  lstm architecture for speech emotion recognition with data augmentation,''
  \emph{arXiv preprint arXiv:1802.05630}, 2018.

\bibitem{speech_aug_gan}
A.~Chatziagapi, G.~Paraskevopoulos, D.~Sgouropoulos, G.~Pantazopoulos,
  M.~Nikandrou, T.~Giannakopoulos, A.~Katsamanis, A.~Potamianos, and
  S.~Narayanan, ``Data augmentation using gans for speech emotion
  recognition.'' in \emph{INTERSPEECH}, 2019, pp. 171--175.

\bibitem{speech_aug_cygan}
F.~Bao, M.~Neumann, and N.~T. Vu, ``Cyclegan-based emotion style transfer as
  data augmentation for speech emotion recognition.'' in \emph{INTERSPEECH},
  2019, pp. 2828--2832.

\bibitem{gan_sahu}
S.~Sahu, R.~Gupta, and C.~Espy-Wilson, ``Modeling feature representations for
  affective speech using generative adversarial networks,'' \emph{IEEE
  Transactions on Affective Computing}, 2020.

\bibitem{window_kelly2011effects}
A.~C. Kelly and C.~Gobl, ``The effects of windowing on the calculation of mfccs
  for different types of speech sounds,'' in \emph{International Conference on
  Nonlinear Speech Processing}.\hskip 1em plus 0.5em minus 0.4em\relax
  Springer, 2011, pp. 111--118.

\bibitem{window_stft_nisar}
S.~Nisar, O.~U. Khan, and M.~Tariq, ``An efficient adaptive window size
  selection method for improving spectrogram visualization,''
  \emph{Computational intelligence and neuroscience}, vol. 2016, 2016.

\bibitem{window_kirkpatrick}
B.~Kirkpatrick, D.~O'Brien, and R.~Scaife, ``A comparison of spectral
  continuity measures as a join cost in concatenative speech synthesis,'' 2006.

\bibitem{chernykh2017emotion}
V.~Chernykh and P.~Prikhodko, ``Emotion recognition from speech with recurrent
  neural networks,'' \emph{arXiv preprint arXiv:1701.08071}, 2017.

\bibitem{tripathi}
S.~Tripathi, T.~Sarthak, and H.~Beigi, ``Multi-modal emotion recognition on
  iemocap dataset using deep learning,'' \emph{arXiv preprint
  arXiv:1804.05788}, 2018.

\bibitem{venkat_Dl_outper_ml}
K.~Venkataramanan and H.~R. Rajamohan, ``Emotion recognition from speech,''
  \emph{arXiv preprint arXiv:1912.10458}, 2019.

\bibitem{iemocap}
C.~Busso, M.~Bulut, C.-C. Lee, A.~Kazemzadeh, E.~Mower, S.~Kim, J.~N. Chang,
  S.~Lee, and S.~S. Narayanan, ``Iemocap: Interactive emotional dyadic motion
  capture database,'' \emph{Language resources and evaluation}, vol.~42, no.~4,
  p. 335, 2008.

\bibitem{ravdess}
S.~R. Livingstone and F.~A. Russo, ``The ryerson audio-visual database of
  emotional speech and song (ravdess): A dynamic, multimodal set of facial and
  vocal expressions in north american english,'' \emph{PLOS ONE}, vol.~13,
  no.~5, pp. 1--35, 05 2018.

\bibitem{savee}
P.~Jackson and S.~ul~haq, ``Surrey audio-visual expressed emotion (savee)
  database,'' 04 2011.

\end{thebibliography}

\end{document}